\documentstyle[12pt]{article}
\documentstyle{fleqn}

\catcode`\@=11
\@addtoreset{equation}{section}

\begin{document}
\begin{titlepage}
\begin{flushright}
OU-HET/192\\
June 1994\\
\end{flushright}

\vspace{5 mm}
\begin{center}
{\large Dynamical Gauge Field induced by the Berry Phase Mechanism}
\vspace{10 mm}

Keiji Kikkawa
\footnote{e-mail address: kikkawa@oskth.kek.ac.jp}
 and
Humitaka Tamura
\footnote{e-mail address: tamu@ssel.toshiba.co.jp}
\\
\vspace{10 mm}
Department of Physics\\
\vspace{1 mm}
Osaka University, Toyonaka, Osaka 560, Japan
\vspace{20 mm}
\end{center}
\begin{abstract}
\ \ \  Some part of the local gauge symmetries in the low energy
region, say, lower than GUT or the Planck energy can be an induced
symmetry describable with the holonomy fields associated with a
topologically non-trivial structure of partially compactified space.

In the case where a six dimensional space is compactified by the
Kaluza-Klein mechanism into a product of the four dimensional
Minkowski space $M_{4}$ and a two dimensional Riemann surface with
the genus $g$, $\Sigma_{g}$, we show that, in a limit where the
compactification mass scale is sent to infinity, a model lagrangian
with a U(1) gauge symmetry produces the dynamical gauge fields in
$M_{4}$ with a product of $g$ U(1)'s symmetry, i.e., U(1)$\times
\cdots\times$U(1). These fields are induced by a Berry phase
mechanism, not by the Kaluza-Klein. The dynamical degrees of freedom
of the induced fields are shown to come from the holonomies, or the
solenoid potentials, associated with the cycles of $\Sigma_{g}$. The
production mechanism of kinetic energy terms for the induced fields
are discussed in detail.
\end{abstract}
\end{titlepage}
\newpage

\baselineskip 25pt
\section{Introduction}

$\qquad$The gauge principle is one of the most fundamental
theoretical
principles in modern physics. In most of theories in particle physics
one usually assumes a high gauge invariance in the Planck or GUT
energy region and then breaks the symmetry into some lower one with
the assumption of the Higgs mechanism. It is therefore considered
natural that the symmetry of system becomes less symmetric as the
energy decreases.

An extreme example of this type can be observed in the Kaluza-Klein
theory\cite{1}. The basic idea of the theory is to begin with a
general diffeomorphism invariance in a higher dimensional
space-time, then one derives some gauge symmetries in the four
dimensional Minkowski space as a subgroup
of the diffeomorphism.

In a previous paper\cite{2}, however, one of the authors pointed out
that an opposite mechanism is possible, namely, the gauge symmetry or
a part of gauge symmetry in low energy physics is able to be
dynamically induced rather than given as a required principle. As a
consequence, the gauge symmetry in low energy region could be higher
than that in high energy region.

The purpose of this paper is to refine the idea of the dynamical
gauge theory proposed in the previous paper and make it clear what
sort of geometrical mechanism is working in the generation of gauge
symmetry.

We begin with a U(1) gauge theory interacting with a fermion and the
gravity in 6-dimensions, then assume that the space is compactified
into the 4-dimensional Minkowski space $M_{4}$ times a 2-dimensional
Riemann surface with genus $g$. We do not ask the detail of the
compactification mechanism but assume that the Kaluza-Klein mechanism
(KKM) has worked there. As a matter of fact a certain number of
components of the metric fields in 6-dimensions are converted into
the gauge potentials in $M_{4}$ by KKM,
but we are not concerned with
them. What we concern is another set of gauge fields converted from
the holonomies whose mechanism will be discussed in detail in the
text.

In many cases the Berry phase potential have attracted physicist's
attention and some have tried to employ them as dynamical fields
\cite{3}, but the difficult feature was how to generate the kinetic
energy terms for them. In some non-linear or four-fermi interaction
models the kinetic energy terms were generated by Feynman graph
summations and the renormalizability was always dubious\cite{4}. In
our discussions therefore a central point is how to generate the
kinetic energy terms for the induced gauge fields.

The dynamical degrees of freedom of the induced gauge fields are
shown to originate in the solenoid potentials\cite{5}, or the
holonomies\cite{6} which are associated
with the independent cycles of $\Sigma_{g}$.
The kinetic energy terms for them are generated from those for the
compactified 2-dimensional components of the vector potential
prepared in the original lagrangian. The mechanism of generating $g$
sets of new kinetic energy terms are discussed in \S 4 and \S 5.

The basic mechanism of the induction is able to be best demonstrated
in the case of the compactified space being $\Sigma_{1}$, i.e., a
torus. In \S 2 we discuss the case of $\Sigma_{1}$ in detail.
Although in this case the induced gauge field decouples from the
matter field, the analysis shows why the gauge field is decoupled and
what sort of modification is needed to get a non-trivial gauge field
coupling with matter.

A simple but nontrivial model is shown to be the case where the
compactified Riemann surface has the genus larger than 1. This is
discussed in \S 5.

The last section is devoted to the discussions and further
outlooks.


\section{A U(1) Model}

$\qquad$We study a model of U(1) gauge theory in 6 dimensions
interacting
with a fermion and gravity, and assume that the space-time is
compactified into a four-dimensional Minkowski space $M_{4}$ times a
torus $\Sigma_{1}$ ( a Riemann surface with genus 1 ). The
compactification is assumed to be caused by the Kaluza-Klein
mechanism. The relevant part of the lagrangian besides the gravity
term is
\begin{equation}
L=\overline{\Psi}i\Gamma^{A}(\partial_{A}+ig^{\prime}V_{A})\Psi-
\frac{1}{4}F_{AB}^{\quad2}+\cdots
\end{equation}
where $A$ and $B$ run from 0 to 5 and $\Gamma$'s are Dirac matrices
defined by
\begin{eqnarray}
\Gamma^{A}&=&\{\Gamma^{\mu}=\gamma^{\mu}\otimes{\bf 1}
\mbox{ for }\mu=0,1,2,3, \mbox{ and }\nonumber \\*
& &\quad\Gamma^{\alpha+3}=\gamma^{5}\otimes i\tau^{\alpha}
\mbox{ for }\alpha=1,2 \qquad\}
\end{eqnarray}
in which the first factors of the direct products refer to the
Minkowski space $M_{4}$ and the second, the Pauli matrices, to the
compactified space $\Sigma_{1}$. A vector in $\Sigma_{1}$ is denoted
with indices $\alpha(=1,2)$, or frequently with a vector notation
like $\vec{\tau}=(\tau^{1},\tau^{2})$. The two periods of the torus
are both taken to be a constant value $l$. The typical size of the
torus is characterized by the Planck length $l\equiv M_{P}^{-1}$. We
further assume that the fermion $\Psi$ stays at the ground state
level of the compactified modes with the wave function $u_{a}(x,
\vec{y})$ ($a=1,2$), hence the full fields are represented as
\begin{eqnarray}
\Psi(x,\vec{y})&=&\frac{1}{l}\psi(x)\otimes u(x,\vec{y}) \\
V_{A}(x,\vec{y})&=&\frac{1}{l}\{V_{\mu}(x),\mbox{ }V_{\alpha}
(x,\vec{y})\}
\end{eqnarray}
where $\psi(x)$ stands for a Dirac spinor in $M_{4}$ and
$\vec{y}$ for the coordinate vector in $\Sigma_{1}$. The constant
$l^{-1}$ is multiplied to adjust the dimension. The wave function
$u$ is normalized as
\begin{equation}
<u^{\dagger}u>\equiv\frac{1}{l^{2}}\int u^{\dagger}u\mbox{ }
d^{2}y=1.
\end{equation}

Substituting (2.3) and (2.4) into (2.1) and integrating over the
compactified space coordinates, one obtains the effective
lagrangian in the Minkowski space
\begin{eqnarray}
{\cal L}&=&<L> \nonumber \\*
&=&\overline{\psi}i\gamma^{\mu}[\partial_{\mu}+igV_{\mu}
+iA_{\mu}]\psi          \nonumber \\*
& &-\frac{1}{4}F_{\mu\nu}^{\quad2}-\frac{1}{2}<F_{\mu\alpha}
^{\quad2}>-\frac{1}{4}<F_{\alpha\beta}^{\quad2}> \nonumber \\*
& &+i\overline{\psi}\gamma^{5}\psi
<u^{\dagger}[i\vec{\tau}\cdot(\vec{\partial}+ig\vec{V})]u>
\end{eqnarray}
where $A_{\mu}$ is the induced potential and defined by
\begin{equation}
A_{\mu}=-i<u^{\dagger}\partial_{\mu}u>
\end{equation}
and $g=g^{\prime}l^{-1}$. The vector $A_{\mu}$ is well-known as
Berry's potential\cite{3}. The lagrangian (2.6) now has an extra
artificial gauge invariance under
$$\begin{array}{lll}
\psi(x)&\rightarrow&\psi(x)e^{i\theta(x)} \\
u(x)   &\rightarrow& e^{-i\theta(x)}u\quad.
\end{array}$$

In $\Sigma_{1}$, we look for such
a classical solution for $\vec{V}$
that provides a constant magnetic field perpendicular to the torus
plane. The solution is
\begin{equation}
\vec{V}_{C}=(By_{2},0).
\end{equation}
Taking account of the small (long wave) fluctuations around
$\vec{V}_{C}$ we choose the potential in $\Sigma_{1}$ as
\begin{equation}
\vec{V}(x,\vec{y})=[v_{1}(x)+y_{2}B]\vec{e}_{1}+v_{2}(x)\vec{e}_{2}.
\end{equation}
For this choice the third and
fourth terms in ${\cal L}$ become $u$-independent and
the equation of motion for $u$ can be derived from the stationary
condition of (2.6) with the normalization (2.5) as
\begin{equation}
Hu\equiv -i\vec{\tau}\cdot\biggl(
\vec{\partial}+ig\vec{V}(x,\vec{y})\biggr)u=\epsilon u,
\end{equation}
where we have assumed that the interaction energy of
$A_\mu$ with $\psi$ is small compared with the Planck energy
as will be discussed below.
The problem is now reduced to the well known Landau motion on a
torus\cite{5} in a uniform magnetic field $B$. As one sees below
(2.10) is exactly solvable and the fermion states are labeled by
the Landau level number $n$.

Before considering the detail we give a comment on the last term in
(2.6). If the eigenvalue $\epsilon$ of (2.10) is happened to be
non-zero, the lagrangian acquires a mass term for $\psi$, because
$i\overline{\psi}\gamma^{5}\psi$ can be converted into $\overline{
\psi}\psi$ by the chiral transformation $\psi\rightarrow\exp(i
\frac{\pi}{4}\gamma^{5})\psi$. The magnitude of the mass must be of
order of the Planck mass because this is the only massive parameter
involved in this model. To get low energy physics only the
possibility is that the system must provide the zero-eigenvalue in
(2.10). This is possible as one see below.

For the potential (2.9), the equation (2.10) is known to have
solutions under the flux quantization (Dirac) condition
\cite{5}\cite{7}
\begin{equation}
\Phi\equiv-Bl^{2}=-\frac{2\pi}{g}N\quad,\quad N=0,1,2,\cdots.
\end{equation}
The eigenvalues are independent of the function $\vec{v}(x)$, which
is called the solenoid potential or the holonomy in mathematics
\cite{5}\cite{6}. The wave function $u(\vec{y})$ on a torus is
defined on its universal covering space.
The uniqueness of wave function on the universal
covering space provides us with the Dirac
condition. The two wave functions $u(\vec{y})$ and $u(\vec{y}
^{\mbox{ }\prime})$ at equivalent positions are usually
different by phase factors, i.e., the solenoid
potentials or the holonomies, which can depend on the external
coordinates $x$ in $M_{4}$. The $x$-dependence of the solenoid
potential plays the crucial role in the next section. The solutions
are given for a fixed $N$ as follows
\cite{5}
\begin{eqnarray}
\epsilon_{\pm n}=\pm\sqrt{2gBn}\quad,n=0,1,2,\cdots \\
u_{0}^{(l)}=&e^{-ig\vec{v}\cdot\vec{y}}\left(\begin{array}{c}
\phi_{0}^{(l)} \\ 0 \end{array}\right)\quad,
\mbox{ for }\epsilon_{0}=0 \\
u_{\pm n}^{(l)}=&\frac{e^{-ig\vec{v}\cdot\vec{y}}}{\sqrt{2}}
\left(\begin{array}{c}
\phi_{n}^{(l)}\\ \mp i\phi_{n-1}^{(l)}\end{array}\right)\quad,
\mbox{ for }\epsilon_{\pm n}\mbox{ }(n\neq0).
\end{eqnarray}
and
\begin{equation}
<u_{n}^{(l)\dagger}u_{n^{\prime}}^{(l^{\prime})}>=
\delta_{nn^{\prime}}\delta_{ll^{\prime}}
\end{equation}
where $\phi_{n}^{(l)}$ ($l=0,1,2,\cdots,N-1$) are $N$ degenerate
eigenfunctions of the Landau problem for particles on a torus, whose
explicit forms are found in ref.\cite{5}.

Before closing this section a comment is in order. As is seen from
the Dirac condition (2.11) and (2.12), the eigenvalue is of order of
the Planck mass and the non-zero eigenstates are not directly related
with the low energy physics in the Minkowski space. Nevertheless, the
excited states with $n\neq0$ play an essential role in generating
the induced fields as seen below.


\section{Local Field Strength}

$\qquad$Let us observe whether the Berry's potential (2.7) has any
dynamical degrees of freedom. In this and the next sections
we assume $N=1$.

Since the $x$-dependence of the spinor function $u(x)$ occurs
through only the solenoid potential $\vec{v}(x)$, we can write
$A_{\mu}(x)$ as
\begin{equation}
A_{\mu}(x)=\partial_{\mu}v^{\alpha}(x)A_{\alpha}(x)
\end{equation}
where
\begin{equation}
A_{\alpha}(x)=-i<u_{0}^{\dagger}\mbox{$\frac{\partial}
{\partial v^{\alpha}}$}u_{0}>
\equiv -i<u_{0}^{\dagger}\partial_\alpha u_0> .
\end{equation}
The induced field strength is then
\begin{eqnarray}
G_{\mu\nu}(x)&=&\partial_{\mu}A_{\nu}(x)
-\partial_{\nu}A_{\mu}(x) \nonumber \\*
&=&\partial_{\mu}v^{\alpha}(x)\partial_{\nu}v^{\beta}(x)
\hat{G}_{\alpha\beta}
\end{eqnarray}
with
\begin{eqnarray}
\hat{G}_{\alpha\beta}&=&\partial_{\alpha}A_{\beta}
-\partial_{\beta}A_{\alpha} \nonumber \\*
&=&-i\{<\partial_{\alpha}u_{0}^{\dagger}\partial_{\beta}u_{0}>-
(\alpha\leftrightarrow\beta)\}.
\end{eqnarray}
The calculation of $\hat{G}_{\alpha\beta}$ is parallel to the Berry
method\cite{8}, i.e., by substituting the complete set in the right
hand side of (3.4), one obtains
\begin{eqnarray}
\hat{G}_{\alpha\beta}&=&-i\sum_{n\neq0}
[<\partial_{\alpha}u_{0}^{\dagger}
u_{n}><u_{n}^{\dagger}\partial_{\beta}u_{0}>-
(\alpha\leftrightarrow\beta)] \nonumber \\*
&=&-i\sum_{n\neq0}\frac{<u_{0}^{\dagger}\partial_{\alpha}H
u_{n}><u_{n}^{\dagger}\partial_{\beta}Hu_{0}>-
(\alpha\leftrightarrow\beta)}{(\epsilon_{0}-\epsilon_{n})^{2}}
\end{eqnarray}
where to get the last equality we have used
\begin{equation}
<u_{m}^{\dagger}\partial_{\alpha}u_{n}>=\frac{<u_{m}^{\dagger}
\partial_{\alpha}Hu_{n}>}{\epsilon_{n}-\epsilon_{m}}
\end{equation}
which is easily derivable by operating $\partial_{\alpha}$ on both
sides of (2.10) and taking the matrix elements. The matrix elements
of the numerator are given as
\begin{equation}
<u_{\pm n}^{\dagger}\partial_{\alpha}Hu_{o}>=
\left\{\begin{array}{ccc}
\pm i\frac{g}{\sqrt{2}}\delta_{n,1}&,&(\alpha=1)\\
\mp \frac{g}{\sqrt{2}}\delta_{n,1}&,&(\alpha=2)
\end{array}\right. \mbox{ for }n>0.
\end{equation}
All transitions to the levels $n\geq2$ are prohibited. Thus we
get the exact result
\begin{equation}
\hat{G}_{12}=-\hat{G}_{21}=\frac{g}{B}.
\end{equation}
The field strength in $x$-space is therefore
\begin{equation}
G_{\mu\nu}=\frac{g}{B}(\partial_{\mu}v_{1}(x)\partial_{\nu}v_{2}(x)-
\partial_{\nu}v_{1}(x)\partial_{\mu}v_{2}(x)).
\end{equation}
For the sake of confirmation we directly calculated the formula (3.4)
with the use of explicit ground state wave function $u_{0}$, and
obtained the same results.

As one sees in (3.9) the gauge field strength $G_{\mu\nu}(x)$ has
two independent local degrees of freedom $v_{1}(x)$ and $v_{2}(x)$,
which are necessary and sufficient for an abelian field. It may be
instructive, at this point, to emphasize the role of the solenoid
potential $\vec{v}(x)$. In the $v$-space the field strength
is simply a constant as in (3.8). The role of the solenoid potential
is to convert the constant induced field in $v$-space to the local
field in $x$-space.
When the vector potential $A_{\mu}(x)$ is line integrated along
a closed curve $C$, $\vec{v}(x)$ draw a closed curve $C^{\prime}$ in
$v$-space on which the constant field $B$ is perpendicularly
being applied. The flux picked up on the $v$-space is the flux in the
$x$-space, which are now $x$-dependent (fig.1).


\section{Effective Lagrangian in 4D}

$\qquad$Now we argue
the generation of the kinetic energy term for the
induced field $G_{\mu\nu}(x)$. For our choice of classical solution
(2.8) and (2.9), the fourth term $<F_{\alpha\beta}^{\quad2}>$ in the
effective lagrangian (2.6) is simply a constant. It makes a
contribution to the cosmological constant, and we disregard it here.

The third term is explicitly written as
\begin{equation}
<F_{\mu\alpha}^{\quad2}>=(\partial_{\mu}v_{1}(x))^{2}+
(\partial_{\mu}v_{2}(x))^{2}.
\end{equation}
If (2.6) is regarded as the lagrangian for the independent field
$\vec{v}(x)$, (4.1) is the kinetic energy term for the scalar fields
$\vec{v}(x)$ and the interactions of them are taken place via
$A_{\mu}$, which is a complicated functional of $\vec{v}$. We now
want to write all of these terms in terms of $A_{\mu}$.

To do this we first suppose that a quantized theory exists with
$\vec{v}$  fields. Then the local product of $v$-fields,
say, such as
\begin{equation}
G_{\mu\nu}^{2}(x)=\frac{g^{2}}{B^{2}}(
\partial_{\mu}v_{1}(x)\partial_{\nu}v_{2}(x)
-\partial_{\nu}v_{1}(x)\partial_{\mu}v_{2}(x))^{2}
\end{equation}
may not be well defined unless any regularization is introduced.
In our case one should remind the theory has a natural cut off
$M_{P}$, the Planck mass, through the Kaluza-Klein mechanism assumed
at the beginning. In our Born-Oppenheimer approximation the four
dimensional coordinates $x^{\mu}$ are regarded as slow parameters,
while the internal coordinates $\vec{y}$ as fast parameters. The
vacuum
expectation values of the local product of two or more operators in
the $x$-space then may be regularized with the Planck mass
parameter. For instance, one can assume for local limit of operator
products
\begin{equation}
\lim_{\mbox{\tiny $\Delta$}\rightarrow l=1/M_{P}}<0|\partial_{\mu}
v^{\alpha}(x)\partial_{\nu}v^{\beta}(x+
\mbox{\scriptsize $\Delta$})|0>
\approx\frac{1}{4}M_{P}^{4}\delta^{\alpha\beta}\eta_{\mu\nu}
\end{equation}
where the Lorentz invariance in 4D space and the rotational
invariance in the compactified space have been assumed, and
$M_{P}^{4}$ comes from the dimensional argument.

If (4.3) is used, the composite operator (4.2) is able to be
decomposed into a sum of normal ordered products,
\begin{eqnarray}
G_{\mu\nu}^{\quad2}(x)&\approx&\frac{g^{2}}{B^{2}}:(
\partial_{\mu}v_{1}\partial_{\nu}v_{2}
-\partial_{\nu}v_{1}\partial_{\mu}v_{2})^{2}: \nonumber \\*
&\mbox{ }&+\frac{3}{2}\frac{M_{P}^{4}}{B^{2}}:
(\partial_{\mu}v_{1})^{2}+(\partial_{\mu}v_{2})^{2}:
+\frac{3}{2}\frac{M_{P}^{8}}{B^{2}}.
\end{eqnarray}
Now, let us take the limit of $M_{P}\rightarrow\infty$
provided that the dimensionless combination $M_{P}^{2}/B\approx
(\frac{2\pi}{g}N)^{-1}$
fixed finite. Then the first term in the r.h.s. of (4.4)
vanishes, and the second term remains finite. The third is a
large constant but should be absorbed into the cosmological
term. Comparing the result with (4.1) we can conclude
\begin{equation}
G_{\mu\nu}^{\quad2}(x)=2e^{2}<F_{\mu\alpha}^{\quad2}>
\end{equation}
where $e$ is a dimensionless constant.

The effective lagrangian of our system in $M_{4}$,
therefore, is expressed as follows
\begin{equation}
{\cal L}=-\frac{1}{4}F_{\mu\nu}^{\quad2}-\frac{1}{4}
G_{\mu\nu}^{\quad2}+\overline{\psi}i\gamma^{\mu}[
\partial_{\mu}+ieA_{\mu}+igV_{\mu}]\psi.
\end{equation}
In this simple model, however, the induced vector field
decouples from the matter. Namely, if one redefines fields as
\begin{eqnarray}
A_{\mu}^{(0)}&=&\frac{1}{\sqrt{e^{2}+g^{2}}}(eA_{\mu}-gV_{\mu})
\nonumber \\*
A_{\mu}^{(1)}&=&\frac{1}{\sqrt{e^{2}+g^{2}}}(gA_{\mu}+eV_{\mu}),
\end{eqnarray}
the lagrangian takes the form
\begin{equation}
{\cal L}=-\frac{1}{4}F_{\mu\nu}^{(0)2}-\frac{1}{4}F_{\mu\nu}
^{(1)2}+\overline{\psi}i\gamma^{\mu}(\partial_{\mu}+ifA_{\mu}^{(1)}
)\psi
\end{equation}
where $f=\sqrt{e^{2}+g^{2}}$.

At this point a couple of comments must be added on (4.8).
The first is  on the number of degrees of
freedom for the induced potential $A_{\mu}$. We have mentioned that
the dynamical degrees of freedom are two if counted in terms of
$\vec{v}$. When the theory have been expressed with $A_{\mu}$, extra
two freedoms are implicitly added. The original theory is therefore
equivalent to the gauge fixed version of the new lagrangian (4.8).
The second is on the Jacobian factor coming from the variable
change $(v_1,v_2) \rightarrow A_\mu$.
The transformation is non-singular at $\vec{v}=0$ because $A_\mu$
vanishes as $\vec{v}$ approaches zero as seen in (3.1).
The term coming from Jacobian therefore may be
expressed in a series of gauge invariant and
local products of various fields
such as
\begin{equation}
a G_{\mu\nu}^2 +
\frac{b}{M_P} G_{\mu\nu} G_{\nu\rho} G_{\rho\mu}+ \cdots,
\end{equation}
where $a$, $b$, $\cdots$ are dimensionless constants. As one
takes a limit of $M_P$ going to infinity, the second term
and those having higher dimensions are negligible.
The first term is able to be renormalized to the
second term in (4.6), hence the effective action
still takes the form of (4.8).

As we mentioned above, however,
the $\Sigma_1$ model provides us with the decoupled gauge
field model as seen in (4.8). The way out of the decoupling is given
in the next section.


\section{Interacting Gauge Models}

$\qquad$In the previous section we derived an induced gauge field,
which
turned out to be decoupled from matter. The model, however, is
instructive because it shows the geometrical mechanism of generating
the local gauge field. It even suggests us how one is able to escape
the decoupling trouble.

Let first observe why it decouples. In the previous model we chose a
single ground state for the internal Hamiltonian (2.10) and adopted
a simple product for the ground state
\begin{equation}
\Psi=\psi(x)u(x,\vec{y}).
\end{equation}
The original theory is invariant under the U(1) gauge
transformation under which
\begin{equation}
\Psi\longrightarrow e^{i\chi}\Psi.
\end{equation}
The artificial gauge transformation associated (5.1) is
\begin{equation}\left\{\begin{array}{lll}
\psi\rightarrow\psi(x)e^{i\theta(x)} \\
u   \rightarrow e^{-i\theta(x)}u(x,\vec{y}).
\end{array}\right.
\end{equation}
Under these situation for the field $\psi$, however, the theory is
invariant even under
$$
\psi\longrightarrow\psi(x)e^{i(\theta+\chi)}
$$
namely, the artificial gauge can be absorbed into the original $U(1)$
gauge transformation. This is the reason of decoupling.

A way out of this problem is therefore to introduce a degenerate set
of ground states for the internal Hamiltonian (2.10). Let
$u^{a}$ ($a=1,2,\cdots,n$) be zero-energy eigenstates of $H$. Then
the ground state should be expressed as
\begin{equation}
\Psi=\sum_{a=1}^{n}\psi^{a}(x)u^{a}(x,\vec{y}).
\end{equation}
This has an $n\times n$ unitary matrix $U(x)$ invariance
\begin{eqnarray}
u(x)&\rightarrow&U(x)u(x) \nonumber \\*
\psi(x)&\rightarrow&\psi(x)U^{-1}(x).
\end{eqnarray}
Then the U(1) part of $U(x)$ is again absorbed into the original
U(1) gauge, but the other SU($n$) parts remain as new degrees of
freedom.

The simplest model might be constructed even for the $\Sigma_{1}$
compactification. Choose the magnetic field $B$ stronger so that the
degeneracy of zero-energy ground states is $N\geq2$. For (5.4) we
are able to introduce $N\times N$ vector potential
\begin{equation}
A_{\mu}^{ab}=-i<u^{\dagger a}\partial_{\mu}u^{b}>.
\end{equation}
We have, however, found that this is essentially equivalent to the
previous model because an explicit construction shows
\begin{equation}
A_{\mu}^{ab}=\delta^{ab}A_{\mu},\quad A_{\mu}^{1}(x)=A_{\mu}^{2}(x)=
\cdots=A_{\mu}^{N}(x) \equiv A_{\mu}(x).
\end{equation}
The reason comes from the fact that, even if one chooses the
magnetic field $B$ stronger, the number of independent solenoid
potentials are still two, hence no way to produce more than single
vector potential.

To get more solenoid potentials, we choose the compactified surface
$\Sigma_{g}$, a Riemann surface with genus $g\geq2$. The independent
number of solenoid potentials, or holonomies, is $2g$ hence $g$
independent vector potentials $A_{\mu}^{(i)}$ are expected.

Next question is whether the massless spinors for (2.10) exist.
For $g\geq2$ if one so chooses the magnetic field $B$ that the
curvature term is canceled, massless spinor solutions do exist. The
Dirac index theorem\cite{9}
$$
{\rm index}\not{\nabla\mbox{ }}=n_{+}-n_{-}
=\frac{1}{2\pi}\int B\neq0
$$
guarantees at least $(n_{+}-n_{-})$ massless stable solutions, where
$n_{\pm}$ represents the number of chiral spinors with chirality
$\pm1$. The number of massless spinors is chosen arbitrarily large
by adjusting the strength $B$ as discussed in ref. \cite{10}.

In the following we prepare $g$ zero-mass spinors
\begin{equation}
u^{a}\qquad(a=1,2,\cdots,g)
\end{equation}
for the Riemann surface of genus $g$. Even if one chooses larger
numbers, the induced gauge potential may not be independent because
the independent number of solenoid potentials are restricted as
discussed above. In fact, even if more than (5.8)
are chosen not all of them are
independent. We choose therefore the $g$ diagonal components
\begin{equation}
A_{\mu}^{a}(x)\equiv-i<u^{\dagger a}\partial_{\mu}u^{a}>
\end{equation}
which are supposed to be independent. The independence of them are
understood by the choice of cycles as a canonical way as shown in
fig.2.

Since, for this choice, the vector potentials (5.9) are all abelian,
the field strengths are presented by the formula as (3.9) for each
superscript $a$.

The crucial point in the generalized
model is how to generate $g$ set
of kinetic energy terms out of $<F_{\mu\alpha}^{\quad2}>$.

Theory of Riemann surfaces\cite{6} tells us that one can always
construct $g$ sets of harmonic 1-form basis $\omega_{i}$ and
anti-harmonic basis $\overline{\omega}_{i}$ which satisfy
\begin{eqnarray}
\int_{a_{i}}\omega_{j}(\vec{y})&=&\delta_{ij} \nonumber \\*
\int_{b_{i}}\omega_{j}(\vec{y})&=&\Omega_{ij}\quad(i,j=1,2,\cdots,g)
\end{eqnarray}
where $(a_{i},b_{i})$ represent canonical
cycles and $\Omega_{ij}$ is
the period $(g\times g)$ matrix, which is symmetric and has positive
imaginary parts.

Now, as the vector potential on $\Sigma_{g}$, we introduce the
following 1-form
\begin{equation}
V(x,\vec{y})=v^{(0)}(\vec{y})+\xi(x,\vec{y})
\end{equation}
where the 1-form potential $v^{(0)}(\vec{y})$ provides a constant
magnetic flux $B$ perpendicular to the surface, and
\begin{equation}
\xi=\sum_{i=1}^{g}v^{(i)}(x)\omega_{i},\quad
\overline{\xi}=\sum_{i=1}^{g}\overline{v^{(i)}}(x)
\overline{\omega_{i}}
\end{equation}
which are curl free in $\vec{y}$ space.

Then the kinetic energy term $<F_{\mu\alpha}^{\quad2}>$
on $\Sigma_{g}$ is given by
\begin{eqnarray}
<F_{\mu\alpha}^{\quad2}>\sim<\partial_{\mu}\overline{\xi},
\partial_{\mu}\xi>=\frac{i}{2}\int\partial_{\mu}\xi\wedge\partial
_{\mu}\overline{\xi} \nonumber \\*
={\rm Im}\Omega_{ij}\partial_{\mu}\overline{v^{(i)}}(x)
\partial_{\mu}v^{(j)}(x).
\end{eqnarray}
The first term $v^{(0)}$ in (5.11) makes no contribution because of
its $x$-independence. Owing to the positivity of the period matrix
${\rm Im}\Omega$, one can diagonalize (5.13) by some linear
transformation and gets
\begin{equation}
<F_{\mu\alpha}^{\quad2}>\sim\sum_{i=1}^{g}[(\partial_{\mu}\tilde{v}_
{1}
^{(i)})^{2}+(\partial_{\mu}\tilde{v}_{2}^{(i)})^{2}]
\end{equation}
where $\tilde{v}$ are linear combinations of $v$'s. The relation
(5.14) guarantees the generation of $g$ set of kinetic energy terms
for the induced potentials (5.9).

As in the case of $\Sigma_{1}$, we have an original U(1) vector
field, and we have generated $g$ set of
abelian vector fields. One of these fields
decouples from the fermions as before, and others couple. After some
orthogonalizations for the gauge potentials we finally obtain the
following lagrangian in the limit of $M_{P}\rightarrow\infty$,
\begin{equation}
{\cal L}=-\sum_{i=0}^{g}(F_{\mu\nu}^{(i)})^{2}+
\sum_{i=1}^{g}\overline{\psi}^{(i)}i\gamma^{\mu}
[\partial_{\mu}+ig_{i}T_{i}V_{\mu}^{(i)}]\psi^{(i)}
\end{equation}
where $T_{i}$ ($i=1,2,\cdots,g-1$) are $g\times g$ matrices which
are traceless, diagonal and mutually orthogonal, and $T_{g}$ is a
unit matrix.

The induced gauge theory has generated a set of new quantum numbers
for fermions which couple with $g$-independent currents.


\section{Comments and Conclusion}

$\qquad$We have demonstrated a new mechanism
of induced gauge theory. We start
off with a higher dimensional space and assume that a part of the
space is compactified in a topologically non-trivial way by the
Kaluza-Klein mechanism. In the limit that the compactification scale
ratio, say, the Planck mass divided by the observable mass scale is
sent to infinity, two kinds of local gauge symmetries are expected.
The first is the well-known Kaluza-Klein gauge field, which is
induced from the compactified space components of the metric tensor
of original space-time.
The second is our gauge fields discussed in the text. The fields are
generated by the local holonomies associated with the cycles of
topologically non-trivial compactified space. The mechanism of gauge
generation is the Berry phase effect. If the quantum states in the
compact space produce vanishing mass for particles in the four
dimensional Minkowski space $M_{4}$, the quantum states in $M_{4}$
are described by the form (5.4). The Berry phase effect then picks
up the $x$-dependent holonomy fields and provides a set of local
gauge fields.

Although we demonstrated an induction of abelian gauge fields, it
will
be straightforward to induce non-abelian gauge if one chooses a
compactified space with non-abelian holonomies. This problem will be
discussed in a future work.

In closing our paper we give some comments.
The first is about a possibility of generating a
new massive scale in the low energy physics. In our arguments
we have disregarded the interactions between
fermion modes ($u$- fields) and assumed that all
energy levels are
degenerated in the compactified space. If the interactions
among $u$-fields are existed, it may be possible to
introduce another scale factor due to the
interaction energy and generate mass splittings
among fermions.

Final comment is about the gauge symmetries of our world. Once people
devoted some time to the study of the Kaluza-Klein theory to
associate
all gauge symmetries to the structure of compactified space. However,
in so far as other mechanisms are shown to be possible for generation
of gauge symmetries, the problem must be reconsidered. One can expect
richer gauge structures from a simpler geometry.

\vskip1.5cm

\noindent Acknowledgments

The authors thank Kiyoshi Higashijima for drawing their
attention to ref.\cite{9} about fermions on surfaces with
higher genus.
This work is supported in part by the Grant-in-Aid for Scientific
Research from the Ministry of Education, Science and Culture
($\sharp06640398$).


\newpage
{\bf Figure Captions}
\begin{description}
\item[Fig.1] \  Mechanism of picking up the local field
$F_{\mu\nu}(x)$.
\item[Fig.2] \ A canonical holonomy basis in $\Sigma_{g}$.
\end{description}

\end{document}